\documentclass[a4paper,pre,superscriptaddress,floatfix,nofootinbib,twocolumn]{revtex4-1}

\usepackage{bbold}
\usepackage{bbm}
\usepackage[pdftex]{graphicx}
\usepackage{latexsym,amsmath,verbatim,amssymb,txfonts, mathtools}
\usepackage{color}
\usepackage{rotating}
\usepackage{verbatim}
\usepackage{multirow}
\usepackage[english]{babel}
\usepackage{comment}
\usepackage{hyperref}
\usepackage{mathrsfs}
\usepackage{cleveref}

\usepackage{todonotes}
\definecolor{dancolor}{rgb}{0.85, 0.1, 1}

\definecolor{jackcolor}{rgb}{0.6, 0.45, 0.15}

\definecolor{filippocolor}{rgb}{0.1, 1.0, 0.85}

\usepackage{tikz}  
\usepackage{qtree,forest}  
\usetikzlibrary{positioning}
\usetikzlibrary{shapes.geometric}
\usetikzlibrary{arrows.meta,arrows}

\begin{document}

\title{Efficient inference of rankings from multi-body comparisons}

\author{Jack Yeung}
\affiliation{Center for Complex Networks and Systems Research, Luddy School
  of Informatics, Computing, and Engineering, Indiana University, Bloomington,
  Indiana 47408, USA}
\author{Daniel Kaiser}
\affiliation{Center for Complex Networks and Systems Research, Luddy School
  of Informatics, Computing, and Engineering, Indiana University, Bloomington,
  Indiana 47408, USA}
\author{Filippo Radicchi}
\affiliation{Center for Complex Networks and Systems Research, Luddy School
  of Informatics, Computing, and Engineering, Indiana University, Bloomington,
  Indiana 47408, USA}
  \email{f.radicchi@gmail.com}

\begin{abstract}
Many of the existing approaches to assess and predict the performance of players, teams or products in competitive contests rely on the assumption that comparisons occur between pairs of such entities.
There are, however, several real contests where more than two entities are part of each comparison, e.g., sports tournaments, multiplayer board and card games, and preference surveys. The Plackett-Luce (PL) model provides a principled approach to infer the ranking of entities involved in such contests characterized by multi-body comparisons. Unfortunately, traditional algorithms used to compute PL rankings suffer from slow convergence limiting the application of the PL model to relatively small-scale systems. We present here an alternative implementation that allows for significant speed-ups and validate its efficiency in both synthetic and real-world sets of data. Further, we perform systematic cross-validation tests concerning the ability of the PL model to predict unobserved comparisons. We find that a PL model trained on a set composed of multi-body comparisons is more predictive than a PL model trained on a set of projected pairwise comparisons derived from the very same training set, emphasizing the need of properly accounting for the true multi-body nature of real-world systems whenever such an information is available.
\end{abstract}

\maketitle

\section*{Introduction}

Competitive contests in the real world often consist of multiple comparisons between pairs of entities, with each comparison characterized by an outcome in favor of one or the other entity. Examples include tournaments in sports like tennis or boxing, and competitions in board games like scrabble or chess. A natural representation for this type of contests is a directed and weighted network or graph~\cite{newman2010networks}. Entities (e.g., players, teams) are represented by the nodes of the graph. Each edge denotes a comparison (e.g., game) between two nodes. The direction of an edge represents a specific outcome of the corresponding comparison, whereas its integer-valued weight quantifies the occurrence of that specific outcome in the contest. There is plenty of existing graph-based approaches that leverage the graph representation of a competitive contest both to assess past performance and to predict future performance of individual entities~\cite{radicchi2011best, motegi2012network, erkol2021best, de2018physical}. 

Among the existing approaches to asses and predict performance in pairwise competitive contests, a special role is certainly played by the Bradley-Terry (BT) model~\cite{bradley1952rank}. This is a principled probability model for the outcome of pairwise comparisons between entities. Given a set of observed comparisons between pairs of entities, the BT model allows to derive scores associated with individual entities. Those scores can be used not just to rank entities, but also to predict outcomes of non-observed comparisons.  The graph representation of the contest is essential to derive properties of the BT model. For example, inferring BT scores is possible through an iterative algorithm whose convergence is guaranteed only if the graph representing the contest is strongly connected~\cite{zermelo1929berechnung}. The BT model has been the subject of many papers and has been applied in the analysis of a large variety of systems~\cite{newman2023efficient, jerdee2023luck}. However, only very recently, it has been shown by Newman that the computational efficiency of inferring the BT scores from observed data can be greatly improved if the equations used in the original formulation by Zermelo are slightly re-arranged~\cite{newman2023efficient}. Typical speed-up factors range between $3$ and $100$ depending on the set of data considered~\cite{newman2023efficient}.

Whereas competitive contests based on pairwise comparisons are abundant, there are also many real competitive contests where more than two entities are involved in each comparison. In football competitions, such as the final tournament of the FIFA World Cup or the UEFA Champions League, there is an initial group-stage round and only the top teams in a group qualify for the following direct elimination rounds~\cite{fifa, uefa}. In a Texas Hold'em Poker tournament, rounds consist of tables composed of several players~\cite{poker}.
Many online video games are played by multiple players simultaneously~\cite{videogames}. It is often the case that multiple products (e.g., songs, movies, books, videos) compete at the same time on a market~\cite{netflix}.  A scientific paper, where the order of the authorship reflects the contribution of each individual to the paper, is often written by more than two authors~\cite{bhattacharya2010authorship}. 

Such systems based on multi-body comparisons are naturally represented by hypergraphs, where hyperedges are vectors denoting the outcome of individual comparisons eventually weighted by the frequency of their occurrence~\cite{plackettAnalysisPermutations1975}. Existing graph-based methods can be adapted to deal with such systems upon prior reduction of the hypergraph to a properly devised graph. However, such an operation is not only responsible for the loss of potentially useful information, but also not specifically tailored for the prediction of the outcome of multi-body comparisons. Based on our knowledge, the Plackett-Luce (PL) model is one of the few principled approaches existing in the literature that can fully account for the ground-truth multi-body structure of the underlying competition. 
The PL model is a probabilistic model that represents the natural extension of the BT model from pairwise to multi-body comparisons. The model was independently introduced by Luce~\cite{luce1958probabilistic} and Plackett~\cite{plackettAnalysisPermutations1975}.
A large literature on the PL model exists, including:
    spectral algorithms to approximate solutions of the PL model
    \cite{maystreFastAccurateInference2015},
    alternative approaches to evaluating the PL model such as a generalized method-of-moments \cite{soufianiGeneralizedMethodofMomentsRank} and other iterative approaches \cite{negahbanIterativeRankingPairwise2012},
    and shortcomings of the model in multimodal rank spaces \cite{seshadriLearningRichRankings2023}.
    Foundational theory on choice theory and decision-making processes and their connection to the PL model has also been conducted; we refer the reader to the classic collection of essays edited by A. A. J. Marley \cite{choiceessays}.
    The PL model is also central in many existing machine learning approaches to so-called learning-to-rank tasks, see for example~\cite{liu2009learning,oosterhuis2022learning,cheng2010label}.

In this paper, we first present the generalization of the computationally efficient algorithm proposed by Newman for the BT model to the PL model. We present a detailed derivation of the iterative algorithm and provide plenty of evidence that such an algorithm displays significant convergence speed-up compared to the usual iteration schema for solving the PL model. Second, we apply the PL model to the analysis of several real-world systems characterized, at least in principle, by multi-body comparisons. These sets of data include: (i) results of national football teams in the FIFA World Cup, (ii) results of football club teams in the UEFA Champions League, (iii) lists of authors of papers in the area of Network Science, (iv) elections of members in the American Psychological Association, (v) two surveys concerning sushi preferences, (vi) a survey concerning the preference for courses at the AGH University of Science and Technology. 
For elections, we find that projecting multi-body comparisons into pairwise comparisons does not lead to any loss in the predictive power of the PL model. The finding poses some doubts on the authenticity of the multi-body nature of this set of data. For the other datasets, we find instead that their multi-body representation is genuine, as the descriptions/predictions based on multi-body  comparisons are generally more accurate than their projected pairwise counter-parts.
Finally, we use the scores inferred by the PL model to generate top-10 rankings of football teams and network scientists.

\section*{Methods}

\subsection*{Ordered and weighted hypergraphs}

We consider systems of $N$ elements/entities where $M$ interactions (e.g., games, rounds, collaborations, comparisons) occur between an arbitrary number $2 \leq K \leq N$ of elements, and the simultaneous presence of interactions of different size $K$ is allowed. We denote with $K_{\min}$ and $K_{\max}$ respectively the minimum and the maximum number of elements involved in a single interaction. These systems can be thought as hypergraphs~\cite{battiston2021physics}, with the peculiarity that their hyperedges are ordered and weighted. An ordered hyperedge is a vector of nodes' labels, i.e., $\vec{\omega} = (\omega_1, \omega_2, \ldots, \omega_K)$, representing for example the rank positions of teams in a group-stage round of a soccer tournament or the list of authors in a scientific paper. The weight $z(\vec{\omega})$ denotes the number of occurrences of the specific hyperedge $\vec{\omega}$, e.g., the number of scientific articles with identical lists of authors. Denoting with $\Omega$ the set of all distinct hyperdeges, we can write $M = \sum_{\vec{\omega} \in \Omega} z(\vec{\omega})$. Note that, if all hyperedges are of size $K=2$, then the corresponding ordered and weighted hypergraph reduces to a directed and weighted graph. 

Differently from standard hypergraphs, two ordered hyperedges involving the very same sets of nodes are considered different if the involved nodes are listed in a different order in the two vectors. Hypergraphs with vector representations of hyperedges were previously considered, among others, in the study of folksonomies~\cite{ghoshal2009random, zlatic2009hypergraph}. There, each hyperedge was composed of exactly $K=3$ components, namely a user, an object and a tag. In our case, however, we do not distinguish types of nodes. Also as we already mentioned, we do not require vectors to have the same number of components. 


\subsection*{Plackett-Luce model}

We consider the so-called Plackett-Luce (PL) model, representing the immediate generalization of the Bradley-Terry (BT) model from directed graphs to ordered hypergraphs
\begin{equation}
    P \left(\vec{\omega} | \vec{\pi} \right) = \prod_{r=1}^{K-1} \, \frac{\pi_{\omega_r}}{\sum_{q=r}^K \, \pi_{\omega_q}} \; .
\label{eq:ho_bt_model}
\end{equation}
Eq.~(\ref{eq:ho_bt_model}) represents the probability of observing the hyperedge  $\vec{\omega}$ given the vector of nodes' scores $\vec{\pi} = \left(\pi_1, \ldots, \pi_i, \ldots, \pi_N \right)$.
Here, $\pi_i = e^{s_i} > 0$ is a numerical value associated with the element $i$ that quantifies the importance/strength of the element $i$ with 
respect to all other elements in the system. 
The scores $\vec{\pi}$ are further subject to the normalization condition
\begin{equation}
\prod_{i=1}^N \pi_i = 1 \; .
    \label{eq:normalization}
\end{equation}
Eq.~(\ref{eq:ho_bt_model}) reduces to the BT model for $K=2$, i.e., $P \left(\omega_1,\omega_2\right) = \pi_{\omega_1} / (\pi_{\omega_1}+\pi_{\omega_2})$. Note that the model of Eq.~(\ref{eq:ho_bt_model}) is invariant to any transformation $\pi_i \to a \pi_i$ for $i = 1, \ldots, N$ and $\forall a > 0$, and the normalization of Eq.~(\ref{eq:normalization})
serves to fix the value of the constant $a$.

The nodes' scores $\vec{\pi}$ must be learned from an observed hypergraph. Given a set of observed hyperedges $\Omega$, the best estimates of the nodes' scores $\vec{\pi}$ are obtained by maximizing the likelihood
\begin{equation}
P\left(\Omega|\vec{\pi} \right) = \prod_{\vec{\omega} \in \Omega} \, \left[ P \left(\vec{\omega} | \vec{\pi} \right) \right]^{z(\vec{\omega})} \; . 
\label{eq:ho_bt_like}
\end{equation}
Such an optimization can be performed by actually maximizing the logarithm of the likelihood, i.e.,
\begin{equation}
\mathcal{L}\left(\Omega|\vec{\pi} \right)  = \log P\left(\Omega|\vec{\pi} \right) =   \sum_{\vec{\omega} \in \Omega} \, z(\vec{\omega}) \, \log \left[ P \left(\vec{\omega} | \vec{\pi}\right) \right] \; .
\label{eq:ho_bt_log_like}
\end{equation}

Indicate with $\Omega^{K,r}_s$ the set of vectors of length $K$ where node $s$ appears at position $r$, i.e., $\Omega^{K,r}_s = \{\vec{\omega} \in \Omega \, | \, \text{dim}(\vec\omega) = K \land \omega_r = s \}$. We can write the portion of the log-likelihood of Eq.~(\ref{eq:ho_bt_log_like}) that contains all terms that depend on the variable $\pi_s$ as

\begin{equation}
\mathcal{L}_s \left(\Omega|\vec{\pi} \right) = \sum_{K=2}^N  \sum_{r=1}^{K} \, \sum_{\vec{\omega} \in \Omega^{K,r}_s} \, z(\vec{\omega})
 \, \left[ \log(\pi_s) -  \sum_{v=1}^r   \log \left( \sum_{i=v}^K \pi_{\omega_i} \right)   \right] 
 \; ,
\label{eq:ho_bt_alt_like}
\end{equation}
We note that $\frac{\partial \mathcal{L}}{\partial \pi_s} \left(\Omega|\vec{\pi} \right) = \frac{\partial \mathcal{L}_s}{\partial \pi_s} \left(\Omega|\vec{\pi} \right)$, thus to maximize the likelihood we take the derivatives of Eq.~(\ref{eq:ho_bt_alt_like}) with respect to $\pi_s$ for each $s=1, \ldots, N$ and set them equal to zero, i.e.,
 \begin{equation}
 \frac{\partial \mathcal{L}_s}{\partial \pi_s} \left(\Omega|\vec{\pi} \right) = 
 \sum_{K=2}^N \, \sum_{r=1}^{K} \, \sum_{\vec{\omega} \in \Omega^{K,r}_s} \, z(\vec{\omega}) 
 \, \left( \frac{1}{\pi_s} - \sum_{v=1}^r   \frac{1}{ \sum_{i=v}^K \pi_{\omega_i}}  \right) = 0
  \;, 
\label{eq:ho_bt_model_log-like_iter}
\end{equation}
 which defines a system of $N$ equations and $N$ unknowns. Such a system of equations has no closed-form solution but can be solved by iteration. 

The most straightforward iterative algorithm is based on
 \begin{equation}
 \pi_s' = \frac{\sum_{K=2}^N \, \sum_{r=1}^{K} \, \sum_{\vec{\omega} \in \Omega^{K,r}_s} \, z(\vec{\omega}) } { \sum_{K=2}^N \, \sum_{r=1}^{K} \, \sum_{\vec{\omega} \in \Omega^{K,r}_s} \, z(\vec{\omega}) 
 \sum_{v=1}^r   1/ \sum_{i=v}^K \pi_{\omega_i}}    \; .
\label{eq:ho_bt_model_iter}
  \end{equation}  
  The above equation represents the generalization 
  to hypergraphs of the iterative algorithm introduced by Zermelo for graphs~\cite{zermelo1929berechnung}.
  Solutions are obtained by starting from some initial vector $\vec{\pi}$ to compute the updated vector $\vec{\pi'}$  using 
  Eq.~(\ref{eq:ho_bt_model_iter}); the vector $\vec{\pi'}$ is then normalized according to Eq.~(\ref{eq:normalization}). These two operations are iterated until convergence. 
  In the iterative algorithm, we consider 
  cyclic asynchronous
  updating, where each score $\pi_s$ is updated in order. One iteration of the algorithm is given by a full update of all $N$ scores $\pi_s$. 
  Asynchronous updates are used to minimize the possibility of the system of equations to enter in a limit cycle; this is a relatively rare situation, and synchronous updating typically leads to the same solution as asynchronous updating.
  At the end of each iteration, we determine the convergence of the algorithm by testing whether the inequality
  \begin{equation}
  A = 
  \sqrt{\frac{1}{N} \, \sum_{s=1}^N \, \left( \frac{\pi_s}{1+\pi_s} -  \frac{\pi'_s}{1+\pi'_s} \right)^2} \leq \epsilon
      \label{eq:convergence}
  \end{equation}
  is satisfied or not. 
  In the above equation, $\frac{\pi_s}{1+\pi_s}$ is the probability that node $s$ beats the average node in a pairwise comparison~\cite{newman2023efficient}. 
  The fact that $1$ is the score of the average player follows directly from the normalization condition of Eq.~(\ref{eq:normalization}).
  We estimate this quantity for each node between consecutive iterations, and quantify $A$ as the mean square of such an error across all elements in the system.  In all our numerical calculations, we use $\epsilon = 10^{-6}$.

  As for the case of the standard BT model on graphs, the iterative procedure of Eq.~(\ref{eq:ho_bt_model_iter}) displays slow convergence, meaning that several iterations are required before the condition of Eq.~(\ref{eq:convergence}) is met. For the BT model, the issue can be addressed by writing a different iterative algorithm as shown by Newman~\cite{newman2023efficient}. We generalize the iterative algorithm by Newman to the PL model as follows. First, we rewrite
  Eq.~(\ref{eq:ho_bt_model_log-like_iter}) as
 \[
 \sum_{K=2}^N \, \sum_{r=1}^{K} \, \sum_{\vec{\omega} \in \Omega^{K,r}_s} \, z(\vec{\omega}) 
 \, \left( \frac{1}{\pi_s} - \frac{1}{ \sum_{i=r}^K \pi_{\omega_i}}  -   \sum_{v=1}^{r-1}   \frac{1}{ \sum_{i=v}^K \pi_{\omega_i}}  \right) = 0 \; .
 \]
 Then, we notice that for any $\vec{\omega} \in \Omega^{K,r}_s$ we have by definition that $\omega_r = s$, thus for each non-null summand in the above equation we can write
 \[
 \frac{1}{\pi_s} - \frac{1}{ \sum_{i=r}^K \pi_{\omega_i}} = \frac {\sum_{i=r}^K \pi_{\omega_i} - \pi_s} {\pi_s \,  \sum_{i=r}^K \pi_{\omega_i} }  = \frac{\sum_{i=r+1}^K \pi_{\omega_i}} {\pi_s \,  \sum_{i=r}^K \pi_{\omega_i} }
 \]
 for $r \leq K-1$. For $r=K$, we automatically have $\frac{1}{\pi_s} - 1/ \sum_{i=r}^K \pi_{\omega_i} = 0$.
Finally, we write the iterative step of the Newman's algorithm for the PL model as
  \begin{equation}
 \pi_s' =  \frac{ \sum_{K=2}^N \, \sum_{r=1}^{K-1} \, \sum_{\vec{\omega} \in \Omega^{K,r}_s} \, z(\vec{\omega}) \,    \, \sum_{i=r+1}^K \pi_{\omega_i} /   \sum_{i=r}^K \pi_{\omega_i}    } {\sum_{K=2}^N \, \sum_{r=1}^{K} \, \sum_{\vec{\omega} \in \Omega^{K,r}_s} \, z(\vec{\omega}) \,  \sum_{v=1}^{r-1}   1/ \sum_{i=v}^K \pi_{\omega_i}} \; .
\label{eq:ho_bt_model_new_iter}
  \end{equation}
  The resulting iterative algorithm requires first the application of Eq.~(\ref{eq:ho_bt_model_new_iter}) and then the imposition of the normalization condition of Eq.~(\ref{eq:normalization}). Still, convergence is assessed using Eq.~(\ref{eq:convergence}).

Both iterative procedures of Eqs.~(\ref{eq:ho_bt_model_iter}) and~(\ref{eq:ho_bt_model_new_iter}) may not converge. The convergence issue is due to the fact that they are obtained by maximizing a likelihood function instead of a posterior probability.
The necessary condition for convergence of the traditional BT model is that the underlying graph is composed of a single strongly connected component~\cite{zermelo1929berechnung}. The same criterion generalizes to the PL model \cite{maystreFastAccurateInference2015}, where the graph that must be strongly connected is the one obtained by properly projecting the hypergraph into its so-called ``full breaking'' graph~\cite{soufianiGeneralizedMethodofMomentsRank}, that is, where every possible directed dyadic edge is included to preserve the strict total ordering implicit in the multi-body comparisons.
To address the convergence issue, we adopt the
same strategy as in Newman~\cite{newman2023efficient}, and assume that the prior for the scores $\vec{\pi}$ is the logistic distribution
\begin{equation}
P(\vec{\pi}) = \prod_{i=1}^N \frac{\pi_i}{(1 + \pi_i)^2} \; ,
\label{eq:prior}
\end{equation}
so that the posterior probability of the model reads
\begin{equation}
P(\vec{\pi} | \Omega) \propto  P(\Omega|\vec{\pi})  \,  P(\vec{\pi}) =  \prod_{\vec{\omega} \in \Omega} \, \left[ P \left(\vec{\omega} | \vec{\pi} \right) \right]^{z(\vec{\omega})} \, \prod_{i=1}^N \frac{\pi_i}{(1 + \pi_i)^2}  \; .
\label{eq:post}
\end{equation}
The maximization of the posterior probability is performed using a similar procedure as the one shown above for the likelihood. The analogue of the Zermelo's iterative step is
\begin{equation}
 \pi_s'  = \frac{1 + \sum_{K=2}^N \, \sum_{r=1}^{K} \, \sum_{\vec{\omega} \in \Omega^{K,r}_s} \, z(\vec{\omega}) } { 2/(\pi_s+1) + \sum_{K=2}^N \, \sum_{r=1}^{K} \, \sum_{\vec{\omega} \in \Omega^{K,r}_s} \, z(\vec{\omega}) 
 \sum_{v=1}^r   1/ \sum_{i=v}^K \pi_{\omega_i}} \; .
\label{eq:ho_bt_model_iter_post}
  \end{equation}
 The Newman's iterative procedure is instead
  \begin{equation}
 \pi_s' =\frac {  1/(\pi_s+1) + \sum_{K=2}^N \, \sum_{r=1}^{K-1} \, \sum_{\vec{\omega} \in \Omega^{K,r}_s} \, z(\vec{\omega}) \sum_{i=r+1}^K \pi_{\omega_i} / \sum_{i=r}^K \pi_{\omega_i}  }  {1/(\pi_s+1) + \sum_{K=2}^N \, \sum_{r=1}^{K} \, \sum_{\vec{\omega} \in \Omega^{K,r}_s} \, z(\vec{\omega}) \,  \, \sum_{v=1}^{r-1} 1/\sum_{i=v}^{K} \pi_{\omega_i}  } \; .
\label{eq:ho_bt_model_new_iter_post}
\end{equation}

As already stressed by Newman, the prior of Eq.~(\ref{eq:prior}) serves as a regularization term that allows the convergence of the iterative scheme irrespective of the connectedness of the underlying hypergraph. Further, imposing the normalization condition of Eq.~(\ref{eq:normalization}) is no longer required. In our application of the method, however, we still apply it after each iteration. We noticed in fact that this helps speeding up the convergence of the iterative algorithm.

\subsection*{Position 1-breaking Plackett-Luce model}

Also, we consider
\begin{equation}
    P \left(\vec{\omega} | \vec{\pi} \right) =  \frac{\pi_{\omega_1}}{\sum_{q=1}^K \, \pi_{\omega_q}} \; ,
\label{eq:ho_bt_model_alt}
\end{equation}
which is a slight modification of the PL model 
of Eq.~(\ref{eq:ho_bt_model}) known in the literature as the
position 1-breaking PL model~\cite{soufianiGeneralizedMethodofMomentsRank}. Note that since $\pi_i = e^{s_i}$, the probability appearing in Eq.~(\ref{eq:ho_bt_model_alt}) is often referred to as the multinomial logistic or softmax function. The model of Eq.~(\ref{eq:ho_bt_model_alt}) gives emphasis to the node occupying the first position in the vector hyperedge
$\vec{\omega}$ only, irrespective of the order in which the other nodes
appear in the hyperedge. 

We can repeat the same exact steps as above to generate an efficient
algorithm able to determine the scores $\vec{\pi}$ of each node in the graph
for a given set of observed hyperedges $\Omega$. The Newman's iterative
equation for this model is
\begin{equation}
 \pi_s' =\frac{ 1/(\pi_s+1) + \sum_{K=2}^N \, \sum_{\vec{\omega} \in \Omega^{K,1}_s} \, z(\vec{\omega}) \, \sum_{i=2}^K \pi_{\omega_i} / \sum_{i=1}^K \pi_{\omega_i}   } { 1/(\pi_s+1) + \sum_{K=2}^N \, \sum_{r=2}^{K} \sum_{\vec{\omega} \in \Omega^{K,r}_s} \, z(\vec{\omega}) /\sum_{i=1}^K \pi_{\omega_i} } \; .
\label{eq:ho_bt_model_new_iter_alt_post}
  \end{equation}

\subsection*{Approximating many-body interactions with pairwise interactions}

An immediate question that can arise is whether the hypergraph-based PL model can be well approximated by properly modified versions of the standard graph-based BT model. An approximation of this type requires generating a graph projection of the hypergraph, by mapping each vector hyperedge to a set of directed edges, and then apply the standard BT model to the graph projection. Please note that the projection should properly account for any eventual weight associated to each hyperedge.

The natural approximation of the model of Eq.~(\ref{eq:ho_bt_model})
is
\begin{equation}
 \tilde{P} \left(\vec{\omega} | \vec{\pi} \right) = \prod_{r=1}^{K-1} \, \prod_{t = r}^{K} \, \frac{\pi_{\omega_r}}{\pi_{\omega_r} + \pi_{\omega_{t}}} \; ,
    \label{eq:ho_bt_model_approx}
\end{equation}
which actually consists in mapping the vector hyperedge
$\vec\omega = \left(\omega_1, \ldots, \omega_K \right)$ to the set
of directed edges $\{(\omega_1, \omega_2), \ldots, (\omega_1, \omega_K), (\omega_2, \omega_3), \ldots, (\omega_2, \omega_K), \ldots, (\omega_{K-1}, \omega_K)\}$.  We refer to the model of Eq.~(\ref{eq:ho_bt_model_approx}) as the projected Plackett-Luce (pPL) model. The above projection of the hypergraph is named ``full breaking'' graph in Ref.~\cite{soufianiGeneralizedMethodofMomentsRank}.

For the model of Eq.~(\ref{eq:ho_bt_model_alt}), we have instead
\begin{equation}
 \tilde{P} \left(\vec{\omega} | \vec{\pi} \right) = \prod_{r=2}^{K}  \, \frac{\pi_{\omega_1}}{\pi_{\omega_1} + \pi_{\omega_{r}}} \; ,
    \label{eq:ho_bt_model_altapprox}
\end{equation}
consisting in mapping the vector hyperedge
$\vec\omega = \left(\omega_1, \ldots, \omega_K \right)$ to the set
of directed edges $\{(\omega_1, \omega_{2}), \ldots,  (\omega_{1}, \omega_{K})\}$. Also in this case, we refer to the model of Eq.~(\ref{eq:ho_bt_model_altapprox}) as the position 1-breaking version of the pPL model. The above projection of the hypergraph is named ``position-1 breaking'' graph in Ref.~\cite{soufianiGeneralizedMethodofMomentsRank}.

\section*{Data}

\begin{table*}[!hbt]
\begin{center}
\scalebox{0.85}{
\begin{tabular}{|l|l|r|r|r|r|r|r|r|}
\hline
Full name & $N$ & $M$ & $K_{\min}$ & $K_{\max}$ & $I$ (Zermelo) & $I$ (Newman) & Speed-up
\\
\hline \hline
Synthetic &   $1000$ & $10000$ & $2$ & $10$ & $103 \pm 1$ & $11.0 \pm 0.1$ & $9$
\\
\hline
Synthetic &   $1000$ & $100000$ & $2$ & $10$ & $169 \pm 1$ & $11.0 \pm 0.1$ & $15$
\\
\hline
FIFA World Cup (FWC) & $83$ & $364$ & $2$ & $4$ & $50 \pm 2$ & $9.0 \pm 0.3$ & $6$
\\
\hline
UEFA Champions League (UCL) & $174$ & $674$ & $2$ & $4$ & $60 \pm 2$ &
$11.1 \pm 0.5$ & $5$
\\
\hline
Sushi 10 & $10$ & $5000$ & $10$ & $10$ & $14 \pm 1$ &
$7.6 \pm 0.5$ & $1.8$
\\
\hline
Sushi 100 & $100$ & $5000$ & $10$ & $10$ & $21 \pm 2$ &
$6.9 \pm 0.3$ & $3$
\\
\hline
AGH course selection 2004 & $7$ & $153$ & $7$ & $7$ & $534 \pm 4$ &
$7.6 \pm 0.4$ & $70$
\\
\hline
APA election 2009 & $10$ & $12078$ & $2$ & $5$ & $16 \pm 1$ &
$7.3 \pm 0.5$ & $2.2$
\\
\hline
Network Science (NS) & $3517$ & $2461$ & $2$ & $14$ & $144 \pm 7$ &
$24 \pm 1$ & $6$
\\
\hline
\end{tabular}
}
\caption{{\bf Summary table.} Each row of the table corresponds to a specific dataset analyzed. From left to right, we report the name of the dataset, the number of nodes $N$, the number of interactions $M$, the minimum $K_{\min}$ and the maximum size $K_{\max}$ of the interactions, the number of iterations $I$ required for convergence for the Zermelo algorithm and for the Newman algorithm, and the average speed-up factor of using the Newman's version of the algorithm instead of the Zermelo's one. The number of iterations have been estimated over at least $10$ runs of the algorithms, each started from a different initial condition. We report average value and standard deviation over those runs.}
\label{tab:1}
\end{center}
\end{table*}

\subsection*{Synthetic hypergraphs}
We consider synthetic ordered and weighted hypergraphs generated as follows. Inputs are the number of nodes $N$, the total number of multi-body interactions $M$, as well as the range $[K_{\min}, K_{\max}]$ for the size of those interactions. We first assign a score $\pi_s$ to each node $s = 1, \ldots, N$ by extracting a random value from the logistic distribution of Eq.~(\ref{eq:prior}). Then, to generate one interaction, we first determine its size $K$ by extracting a random number at uniform from the interval $[K_{\min}, K_{\max}]$. Then, we extract, at uniform and without replacement, $K$ nodes out of the $N$ available; those nodes are ordered at random according to either Eq.~(\ref{eq:ho_bt_model}) or Eq.~(\ref{eq:ho_bt_model_alt}) to generate the actual hyperedge $\vec{\omega}$; while building the network, we keep track of the weight 
$z(\vec{\omega})$ associated with each hyperedge $\vec{\omega}$.

\subsection*{FIFA World Cup}

We analyze data concerning the outcome of the
different rounds played by football national teams 
in all $22$ FIFA World Cup (FWC) final 
tournaments held between $1930$ and $2022$.
Data are retrieved from the Wikipedia pages of the individual tournaments, 
see for example \url{https://en.wikipedia.org/wiki/1982_FIFA_World_Cup}
for data concerning the FWC tournament of Spain $1982$.
The dataset consists of $N = 83$ total teams, with the caveat
that we merge both ``West Germany'' and ``East Germany''
into the single team ``Germany'' between $1954$ and $1990$.
The dataset is composed of $364$ total rounds of rather
different type, given that the
format of tournament changed dramatically in the course
of the years. The format of the most recent tournaments 
consists of an initial round
played in groups of four teams with the first two teams of each group
proceeding to the following direct elimination rounds. There were, however,
tournaments consisting of only direct eliminations rounds (e.g., Italy $1934$) and with more than one group-stage round (e.g., Spain $1982$).
Each round is considered as an ordered hyperedge in our analysis.
The dataset contains $M = 364$ hyperedges, with size ranging between $K_{\min} = 2$ and $K_{\max} = 4$.

\subsection*{UEFA Champions League}

We consider data reporting the outcome of all rounds played 
in the UEFA Champions League (UCL) tournaments between seasons $1992-1993$ and $2023-2024$.
Preliminary/qualification rounds are excluded.  
Our sources of data are the Wikipedia pages of the individual editions, see
for example \url{https://en.wikipedia.org/wiki/2022–23_UEFA_Champions_League} for data concerning the UCL tournament of season $2022-2023$.
The resulting dataset consists of $N = 174$ teams and $M = 674$ rounds of size between $K_{\min}=2$ and $K_{\max}=4$.
Also in this case, the tournament structure changed drastically in the course of the years. For example, the $1992-1993$ tournament consisted of two direct elimination rounds, a round with groups of size four, 
and a final direct elimination round. In $2002-2003$, there were two rounds with groups of size four, and then direct elimination rounds. In the most recent tournaments, an initial round with groups of four teams was followed by direct elimination rounds. In UCL tournaments, two teams generally played one against the other two times, with the exception of the final that was played in a single match. Also here, the outcome of each round is represented as an ordered hyperedge in the hypergraph-based PL model.


\subsection*{Sushi preferences}

We consider results of a series of surveys conducted by T. Kamishima asking $M = 5000$ individuals for their preferences about various kinds of sushi~\cite{kamishima2003nantonac}. We consider two distinct datasets: one containing $K_{\min} = K_{\max} =10$ complete strict rank orders of $N = 10$ different kinds of sushi; the other 
concerning $K_{\min} = K_{\max} =10$ strict rank orderings for $N = 100$ different kinds of sushi. Data are obtained from \url{https://preflib.simonrey.fr/dataset/00014}.

\subsection*{AGH course selections}

We consider a dataset containing the results of a survey for $M = 153$ students at the AGH University of Science and Technology (\url{https://www.agh.edu.pl/en/}) about their course preferences for year $2004$~\cite{skowron2013achieving}. Each student provided a rank ordering over all $N = K_{\min} = K_{\max} = 7$  courses.  Our source of data is \url{https://preflib.simonrey.fr/dataset/00009}.

\subsection*{APA elections}

We consider a dataset containing the results of the elections of the American Psychological Association (APA, \url{https://www.apa.org})  in year $2009$~\cite{regenwetter2007unexpected}. The voters were allowed to rank any number up to $K_{\max} =5$ candidates without ties. We exclude data where only one candidate is listed, thus imposing $K_{\min} =2$. The total number of voters in the dataset is $M = 12078$.  Data are collected from \url{https://preflib.simonrey.fr/dataset/00028}.

\subsection*{Network Science collaborations}

Data are from the November 2024 version of OpenAlex (\url{openalex.org}).
We consider all papers published between $1999$ and $2023$
in the following journals: Nature, Nature Communications, Nature Physics, Proceedings of the National Academy of Sciences USA, Physical Review E, Physical Review Letters, Physical Review X, Science and Science Advances.
Only papers with topic ``Statistical Mechanics of Complex Networks'' (topic id T10064 in the OpenAlex database) are selected. For each paper, we retrieve the ordered list of authors. We rely on the authors' OpenAlex identifiers for disambiguation. We exclude single-author papers as they do generate any hyperedge in our hypergraph. The resulting dataset consists of $M=2461$ papers authored by $N=3517$ distinct scholars; the size of the teams involved in the writing of the papers range between $K_{\min}=2$ and $K_{\max}=14$. Our main goal in considering this dataset is quantifying the leadership of individual scientists, typically reflected in the order of authorship in this specific scientific discipline (i.e., the senior author appears in the last position in the list of authors). To make the dataset consistent with the others considered in this paper, we actually invert the order of the authors in each paper, so that the first authors appears in the last position in the corresponding hyperedge, and the last author appears as the first element in the corresponding hyperedge.

\section*{Results}

\subsection*{Convergence of the algorithms}

\begin{figure}[!hbt]
    \centering
    \includegraphics[width=0.95\columnwidth]{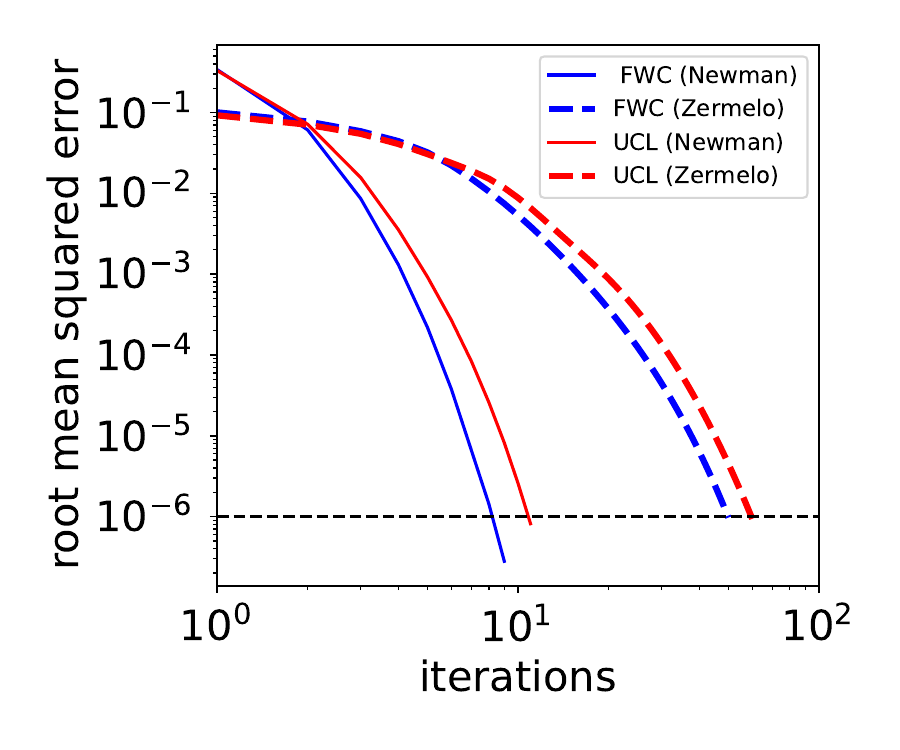}
    \caption{{\bf Convergence of the algorithms.} 
    We plot the root mean squared error of Eq.~(\ref{eq:convergence})
    as a function of the number of iterations for the schemas of Eqs.~(\ref{eq:ho_bt_model_iter_post}) [dashed curves] and~(\ref{eq:ho_bt_model_new_iter_post}) [full curves]. We display results for the FIFA World Cup (FWC) [blue] and the UEFA Champions League (UCL) [red] datasets. The horizontal black dashed line denotes $\epsilon = 10^{-6}$, i.e., the accuracy parameter of Eq.~(\ref{eq:convergence}) used to determine the convergence of the iterative algorithms.
    }
    \label{fig:1}
\end{figure}

First of all, we report on some tests aimed at demonstrating the 
improved efficiency of the iteration algorithm 
{\it \'a la} Newman of Eq.~(\ref{eq:ho_bt_model_new_iter_post}) compared to the standard Zermelo's schema of Eq.~(\ref{eq:ho_bt_model_iter_post}). 
In all our tests, we ensure that the initial guess of the vector $\vec{\pi}$ is the same for both the iterative schemes; also, we use the same exact criterion of
Eq.~(\ref{eq:convergence}) to estimate the convergence of each algorithm. We then compare the total number of iterations required by either one method or the other to converge. Results of this analysis for a selection of two real datasets are displayed in Figure~\ref{fig:1} and summarized in Table~\ref{tab:1} for all sets of real and synthetic data. Speed-ups in convergence are apparent: depending on the dataset considered, we roughly observe a $5$ to $400$ speed-up factor.

\subsection*{Prediction of unobserved events}

\begin{figure*}[!hbt]
    \centering
    \includegraphics[width=0.95\textwidth]{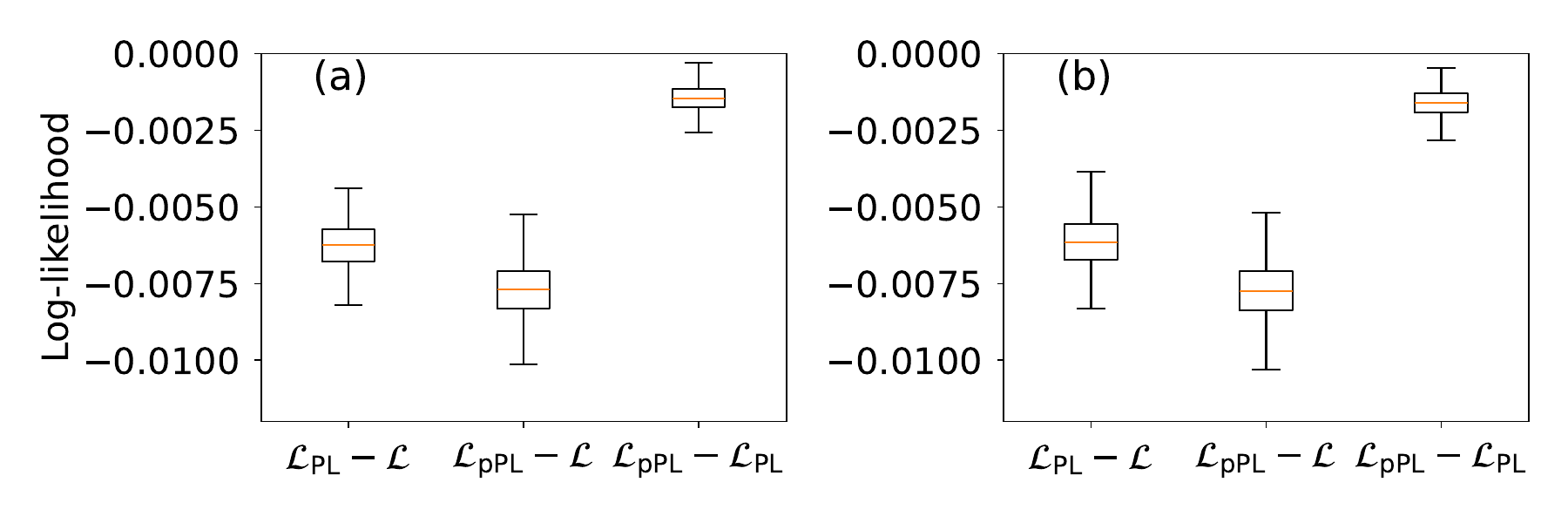}
    \caption{{\bf Model selection for synthetic hypergraphs.}
    (a) We consider synthetic hypergraphs with $N=1000$ nodes and $M=100000$ interactions. The size of those interactions ranges between $K_{\min} =2$ and $K_{\max} =10$ nodes. Nodes' scores are preassigned from the model of Eq.~(\ref{eq:post}) and  the order of the nodes in each  interaction is determined on the basis of Eq.~(\ref{eq:ho_bt_model}). We train the Plackett-Luce (PL) and its projected version (pPL) using $80\%$ of the interactions, and make predictions on the remaining $20\%$. We estimate the logarithm of the likelihoods of the two models, respectively $\mathcal{L}_{\text{PL}}$ and $\mathcal{L}_{\text{pPL}}$, using Eq.~(\ref{eq:ho_bt_log_like}); at the same time, we also estimate the logarithm of the likelihood $\mathcal{L}$ of the unobserved interactions using the ground-truth value of the nodes' scores. We display results over $1000$ independent realizations of the prediction task; the orange line identifies the median, the boxes delimit the the first and third quartiles, and the error bars denote the $5\%$ and $95\%$ confidence intervals. (b) Same as in panel (a), but for the position 1-breaking PL model as defined in Eq.~(\ref{eq:ho_bt_model_alt}).
    }
    \label{fig:2}
\end{figure*}

Next, we compare the PL model of Eq.~(\ref{eq:ho_bt_model}) against its graph counterpart pPL of Eq.~(\ref{eq:ho_bt_model_approx}). We do not show any results here, but we systematically verify that the two approaches infer vectors of scores $\vec{\pi}$ whose components are highly correlated. In Figure~\ref{fig:2}(a), we show results concerning the predictive power of the two models when applied to synthetic hypergraphs generated according to the PL model of Eq.~(\ref{eq:ho_bt_model}). We specifically focus on cross-validation tests where $80\%$ of the interactions are used to train the models, and the unobserved $20\%$ are used to quantify their predictive powers. For both models, predictive performance is quantified using the log-likelihood of Eq.~(\ref{eq:ho_bt_log_like}) estimated for the unobserved $20\%$ data; we denote this quantity as $\mathcal{L}_{\text{PL}}$ for the PL model and as $\mathcal{L}_{\text{pPL}}$ for the pPL model; we also estimate the log-likelihood $\mathcal{L}$ using the ground-truth scores $\vec{\pi}$ used to generate the synthetic hypergraph. The results of Figure~\ref{fig:2}(a) clearly show that the true model consistently outperforms its approximated version.
The same conclusions are valid also for the position 1-breaking versions of PL and pPL, respectively obtained using 
Eqs.~(\ref{eq:ho_bt_model_alt}) and~(\ref{eq:ho_bt_model_altapprox}), see Figure~\ref{fig:2}(b).

\begin{figure*}[!hbt]
    \centering
    \includegraphics[width=0.95\textwidth]{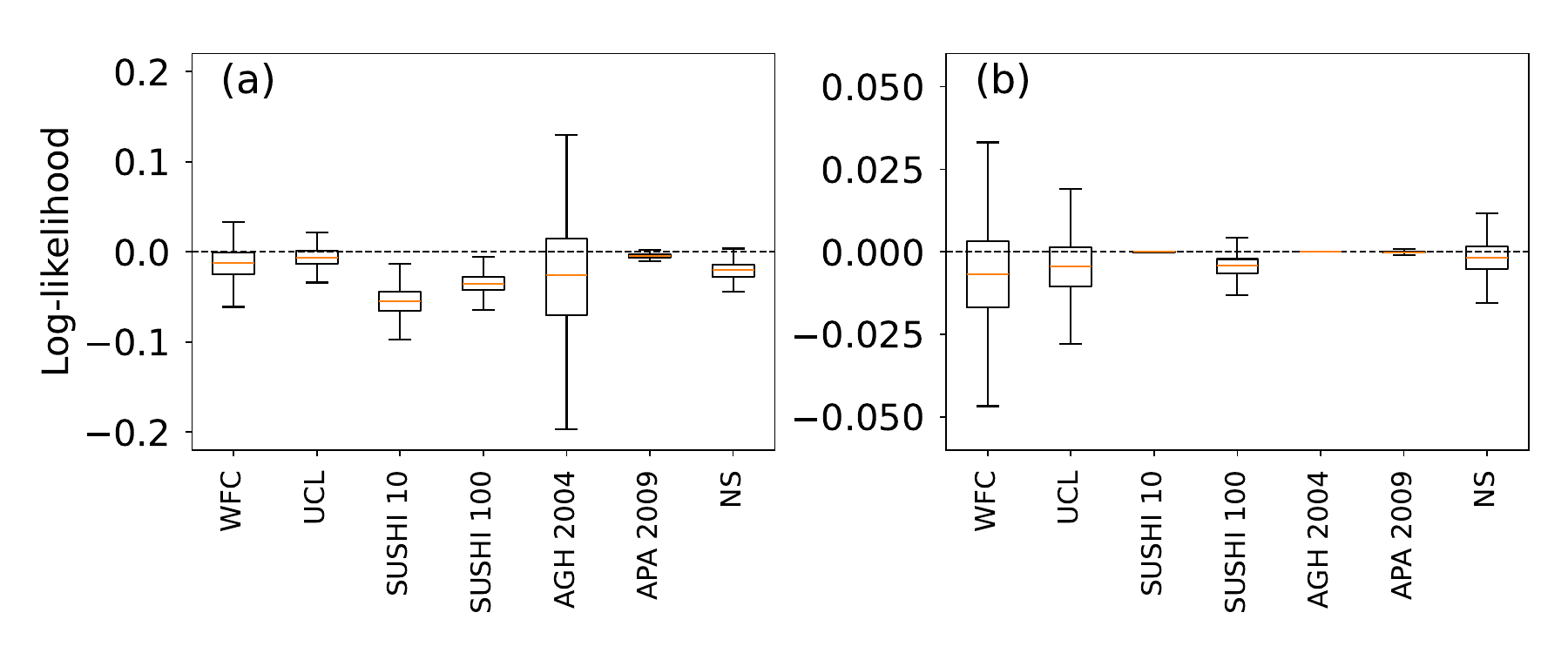}
    \caption{{\bf Model selection for real hypergraphs.}
    (a) Same as in Figure~\ref{fig:2}(a), but for the real datasets considered in the paper. Each plot summarizes the statistics of the log-likelihood difference $\mathcal{L}_{\text{pPL}} - \mathcal{L}_{\text{PL}}$ over $1000$ realizations of the prediction task where $80\%$ of the interactions are used to infer the nodes' scores then used to predict the remaining $20\%$ of unobserved interactions. (b) Same as in (a), but for the position 1-breaking PL model.
    }
    \label{fig:3}
\end{figure*}

We repeat the same cross-validation tests also for the real-world datasets. Results are summarized in Figure~\ref{fig:3}(a) for the standard PL model, and in Figure~\ref{fig:3}(b) for its position 1-breaking variant. For real-word datasets, two fundamental observations are in order. First, the actual score of the individual nodes is not known. This fact implies that predictions can not be judged in an absolute fashion rather only relatively to other predictive models. Second and more important, although all real systems are represented as ordered and weighted hypergraphs, there is no guarantee that this is their {\it bona fide} representation. The very comparison of the predictive power of the multi-body PL against the pairwise pPL actually offers the possibility of stating whether interactions are better represented by multi-body comparisons rather than the simpler aggregation of pairwise comparisons. Based on the results of Figure~\ref{fig:3}(a), we conclude that the multi-body-comparison hypothesis is meaningful for all real datasets considered in this paper; the only exception is represented by the APA 2009 dataset, where most of the interactions can be equally well explained as being the aggregation of multiple pairwise comparisons. The full version of the position 1-breaking PL model appears instead at least as predictive as its projection-based variant for all datasets considered, see Figure~\ref{fig:3}(b). For WFC and UCL, we still observe a better performance of the hypergraph-based approach: this is probably due to the fact that the vast majority of the comparisons is between pairs of elements, thus no great difference between the PL and the position 1-breaking PL models is expected. For the NS data also, we observe comparable performance denoting that the full order of the authors is reflecting well their contribution in the collaboration. For survey/election data instead, the result seems to indicate that respondents/voters identify the first element by comparing it against each of the other alternatives in pairwise comparisons. 

\subsection*{Ranking of football teams and network scientists}

We conclude our journey by providing ranked lists of football teams and network scientists using the data at our disposal. In all cases, we generate those lists using the entire datasets at our disposal to infer the nodes' scores $\vec{\pi}$. For FWC and UCL, we use the PL model of Eq.~(\ref{eq:ho_bt_model}). For NS, we rely on the position 1-breaking PL model of Eq.~(\ref{eq:ho_bt_model_alt}). For brevity, we list in Table~\ref{tab:2} only the top-10 elements in the rankings; complete rankings are available on the github repository 
\url{https://github.com/jackyeung99/higher_order_ranking}.

\begin{table}[!hbt]
\begin{tabular}{|r|l|l|l|}
\hline
Rank & FWC & UCL & NS 
\\
\hline \hline
1 & Brazil & Real Madrid & A. Vespignani
\\
\hline
2 & Germany & Bayern Munich & H.A. Makse
\\
\hline
3 & Italy & Barcelona & L.A.N. Amaral
\\
\hline
4 & Argentina & Liverpool & R. Pastor‐Satorras
\\
\hline
5 & Netherlands & Chelsea & K. Sneppen
\\
\hline
6 & France & Manchester City & A.‐L. Barab\'asi
\\
\hline
7 & Croatia & Juventus & L. Zdeborov\'a
\\
\hline
8 & England & Milan & S. Havlin
\\
\hline
9 & Sweden & Paris Saint-Germain & J.F.F. Mendes
\\
\hline
10 & Czechoslovakia & Atl\'etico Madrid & H.E. Stanley
\\
\hline
\end{tabular}
\caption{{\bf Top-$10$ ranking in football and network science.} We display the items ranking in top-$10$ positions in the datasets FIFA World Cup (WFC), UEFA Champions League (UCL), and the Network Science (NS).
}
\label{tab:2}
\end{table}

In the WFC dataset, we see that the first four positions are occupied by Brazil, Germany, Italy and Argentina, i.e., the most successful national teams in the history of the World Cup. All others are also good examples of teams characterized by historically good performance in the competition. The fact that Netherlands, Croatia, Sweden and Czechoslovakia are in the top $10$ even if they did not win any World Cup so far is somehow surprising. 

Similar conclusions are valid also for the UCL dataset, where Real Madrid dominates the ranking followed by Bayern Munich, Barcelona and Liverpool. It is a bit unexpected that Milan appears only at the 8th position in the table in spite of being the second most successful team in the competition. We stress, however, that our dataset covers only the period $1993-2024$, when Milan won $3$ UCL titles.  


For the NS dataset, we first rank scholars using the model of Eq.~(\ref{eq:ho_bt_model_alt}); we then consider only scholars with more than $10$ papers in our set of data. A. Vespignani tops the ranking being the senior author in all the $28$ papers he co-authored according to our dataset. All other scholars in the top-10 table have very high percentages of papers where they appear as senior authors combined with an overall large number of publications. Specifically, we have that: H.A. Makse has $13$ papers with $92\%$ of senior-author positions, L.A.N. Amaral $17$ and $88\%$, R. Pastor-Satorras $49$ and $67\%$, K. Sneppen $11$ and $91\%$, A.L. Barab\'asi $34$ and $76\%$, L. Zdeborov\'a $11$ and $91\%$,  S. Havlin $95$ and $56\%$, J.F.F. Mendes $53$ and $77\%$, and H.E. Stanley $70$ and $50\%$. We stress that being the senior author in  many papers is not the only important aspect to reach a top rank position; a very important factor is also to be the senior author in papers with other authors with large scores. At the same time, we remark that the contribution of being the senior author of a paper to the score of a scholar is proportional to the number of authors in the paper. Also, scholars with many single-author publications are penalized as these publications are not accounted for in this specific ranking.

\section*{Conclusions}
We successfully generalized a recent iteration schema for ranking entities from pairwise comparisons to the multi-body-comparison setting. Alongside a formal derivation of the iterative algorithm, we provided numerical evidence that the proposed method converges faster than the state-of-the-art iterative procedure in a variety of synthetic and real-world datasets.

We then performed a systematic analysis aimed at predicting the outcome of unobserved comparisons in various real-world systems. 
For all datasets considered, models that take advantage of the multi-body nature of the comparisons display larger predictive power than that of models based on pairwise projections of such multi-body comparisons. In election data, both approaches have comparable predictive power, questioning the {\it bona fide} multi-body nature of this dataset.

Data and code necessary to reproduce our analyses are available at 
\url{https://github.com/jackyeung99/higher_order_ranking}.

\section*{Acknowledgements}

The authors thank G. Bianconi for comments and suggestions on the paper.
This work received partial support from the Air Force Office of Scientific
Research (grant numbers FA9550-21-1-0446 and FA9550-24-1-0039). The funders had no role in study design, data collection, and analysis, the decision to publish, or any opinions, findings, conclusions, or recommendations expressed in the manuscript.

\bibliography{bibliography}

\end{document}